\documentstyle[aps,multicol,prl,epsfig]{revtex}

\newcommand{\beq}{\begin{equation}}
\newcommand{\eeq}{\end{equation}}

\begin{document}

\begin{title}
{\bf Averaging For Solitons With Nonlinearity Management}
\end{title}

\author{D.E. Pelinovsky$^1$, P.G. Kevrekidis$^{2}$ and D.J. Frantzeskakis$^{3}$}
\address{
$^{1}$ Department of Mathematics, McMaster University, Hamilton,
Ontario, Canada, L8S 4K1 \\ $^{2}$ Department of Mathematics and
Statistics, University of Massachusetts, Amherst MA 01003-4515, USA \\
$^{3}$ Department of Physics, University of Athens,
Panepistimiopolis, Zografos, Athens 15784, Greece}

\date{\today}
\maketitle

\begin{abstract}
We develop an averaging method for solitons of the nonlinear Schr{\"o}dinger
equation with periodically varying nonlinearity coefficient. This
method is used to effectively describe solitons in Bose-Einstein
condensates, in the context of the recently proposed and experimentally
realizable technique of Feshbach resonance management. Using the derived
local averaged equation, we study matter-wave bright and dark solitons and
demonstrate a very good agreement between solutions of the averaged and
full equations.
\end{abstract}

\vspace{3mm}

\begin{multicols}{2}

{\it Introduction}. Dispersive nonlinear wave equations are
appropriate mathematical models for various nonlinear phenomena in
fluid mechanics, optics, plasmas and condensed matter physics. The
prototypical equation that generically emerges in the description of
envelope waves is the nonlinear Schr{\"o}dinger (NLS) equation
\cite{sulem,sulem1,sulem2} of the form:
\begin{eqnarray}
i u_t=-D \Delta u + \Gamma |u|^2 u + V({\bf x}) u.
\label{eqn1}
\end{eqnarray}
Here $u({\bf x},t)$ is a complex envelope field, $V({\bf x})$ is
an external potential, $\Delta$ is the Laplacian operator in multi
dimensions, and $D$ and $\Gamma$ are the coefficients of the
dispersive and nonlinear terms respectively.

In a number of physical applications, the coefficients $D$ and
$\Gamma$ exhibit temporal periodic variations. When $D = D(t)$,
the NLS equation (\ref{eqn1}) describes the dispersion management
(DM) scheme in fiber optics, which is based on periodic
alternation of fibers with opposite signs of the group-velocity
dispersion. The DM scheme supports robust breathing solitons
\cite{DMS}, which are well described through the averaging method
by the integral NLS equation \cite{DMS1}. Extensions of the
averaging method were developed for {\em strong} management with
large variations of the dispersion coefficient \cite{PZ} and for
{\em weak} management with small variations of the dispersion
coefficient \cite{YK}.

When $\Gamma = \Gamma(t)$, the NLS equation (\ref{eqn1}) has
applications in optics for transverse beam propagation in layered
optical media \cite{mal1}, as well as in atomic physics for the
Feshbach resonance \cite{inouye} of the scattering length of
inter-atomic interactions in Bose-Einstein condensates (BECs). The
periodic variation of the scattering length by means of an
external magnetic field provides an experimentally realizable
protocol for the generation of robust matter-wave breathers
\cite{KTFM}, and for their persistence against collapse type
phenomena in higher dimensions \cite{abdul,saito}. Solitary waves
have become a focal point in studies of BEC both theoretically and
experimentally \cite{solitons,herrero} due to their coherence
properties. Hence, nonlinearity management using Feshbach
resonance promises to provide a viable alternative for the
generation of coherent nonlinear wave structures.

Given the importance in nonlinear optics and condensed matter physics,
of applications of the NLS equation
(\ref{eqn1}) with periodically varying nonlinearity
coefficient, we
extend the averaging method of \cite{DMS1,PZ} to solitons with {\em
strong} nonlinearity management, when the periodic variations of the
nonlinearity coefficient are large in amplitude. Comparing with 
earlier works, we note
that the averaged equation for strong dispersion management in
\cite{DMS1,PZ} is {\it nonlocal}, whereas our main averaged
equation (see Eq. (\ref{GP-main})) for strong nonlinearity
management is {\it local}. Furthermore, our averaging method is
more general than the asymptotic expansion method, exploited for
weak dispersion management in \cite{YK} and for weak
nonlinearity management in \cite{abdul}. Since the averaged
equation obtained herein is simple, we compute numerically 
solitary waves of the averaged equation and compare with
those of the full problem, showing the excellent agreement
between the two.

We emphasize that the main contribution of this work
is two-fold. From a mathematical point of view, 
it is the derivation of a novel averaged equation 
that describes dynamics of solitary waves 
under nonlinearity management. From a physical point 
of view, the main result is the computation of the 
parameter domains, where nonlinear waves exist
in the BEC and nonlinear optics. We hope that highlighting
the relevant analogies and differences may
also stimulate additional cross-fertilization between
these sub-disciplines and their respective mathematical
techniques.
Furthermore, our work 
poses the interesting problem of understanding 
what happens in the parameter domains, where 
solitary waves do not exist. These fundamental 
problems are of interest not only to 
theorists and experimentalists in atomic and optical 
physics but also, more generally, to researchers in nonlinear 
and wave physics, 
where periodic temporal variations and
their averaging methods are studied 
(see e.g., \cite{sanders}).

{\it Derivation of the averaged equation}. We start with the NLS
equation (\ref{eqn1}) with $D = 1$ and $\Gamma = \Gamma(t)$. The
external potential $V(x)$ is left arbitrary but we keep in mind
that the magnetic and laser trappings relevant to BEC applications
impose parabolic and periodic potentials respectively. Also, we
will restrict ourselves to one spatial dimension, but
generalization of the method to multi-dimensions is
straightforward. The resulting equation (also referred to as the
Gross--Pitaevskii (GP) equation \cite{sulem2}) describes the
``cigar-shaped'' BECs and reads:
\begin{equation}
\label{GP} i u_t = - u_{xx} + \Gamma(t) |u|^2 u + V(x) u,
\end{equation}
where the nonlinearity coefficient (proportional to the scattering
length in the BECs) $\Gamma(t+\epsilon) = \Gamma(t)$ is a smooth,
sign-indefinite, periodic function of period $\epsilon$. We assume
that the period $\epsilon$ of the nonlinearity management is small
compared to the characteristic propagation time of nonlinear
waves, while the nonlinearity variations are large in amplitude.
In this case, we decompose $\Gamma(t)$ into the mean-value part
$\gamma_0$ and a large fast-varying part $\gamma$, according to
the representation:
\begin{equation}
\label{gamma} \Gamma(t) = \gamma_0 + \frac{1}{\epsilon}
\gamma(\tau), \qquad \tau =  \frac{t}{\epsilon},
\end{equation}
where $\gamma(\tau + 1) = \gamma(\tau)$ and $\int_0^1 \gamma(\tau)
d \tau = 0$. Using
\begin{equation}
\label{transformation} u(x,t) = v(x,\tau) \exp \left( - i
\int_0^{\tau} \gamma(\tau') |v|^2(x,\tau') d\tau' \right),
\end{equation}
we remove the large fast variations of the nonlinearity
coefficient and bring the GP equation (\ref{GP}) to an equivalent
form:
\begin{eqnarray}
\nonumber && i \epsilon^{-1} v_{\tau}   =  - v_{xx} + \gamma_0
|v|^2 v + V(x) v \\
\nonumber && + 2 i v_x \int_0^{\tau} \gamma(\tau')
|v|^2_{x}(x,\tau') d\tau' +  i v \int_0^{\tau} \gamma(\tau')
|v|^2_{xx}(x,\tau') d\tau' \\
\label{GP-equivalent} && + v \left( \int_0^{\tau} \gamma(\tau')
|v|^2_{x}(x,\tau') d\tau' \right)^2.
\end{eqnarray}

In the averaging method (see \cite{PZ} for details), we decompose
solutions of the problem with variable coefficients
(\ref{GP-equivalent}) into a slowly varying mean part $w(x,t)$ and
a small, fast-varying part $v_1(x,\tau)$:
\begin{equation}
\label{decomposition} v(x,\tau) = w(x,t) + \epsilon
v_1(x,\tau;w(x,t)), \qquad t = \epsilon \tau.
\end{equation}
The varying part $v_1(x,\tau;w)$ is a periodic function of $\tau$
with unit period. To leading order, this condition is
satisfied if $w(x,t)$ satisfies the averaged equation:
\begin{eqnarray}
\label{GP-averaged}
\nonumber
i w_t  = & - & w_{xx} + \gamma_0 |w|^2 w + V(x) w  \\
& + & 2 i \nu_1 w_x |w|^2_{x} + i \nu_1 w |w|^2_{xx} + \nu_2 w
\left( |w|^2_{x} \right)^2,
\end{eqnarray}
where $\nu_1 = \int_0^1 \nu(\tau) d \tau$, $\nu_2 = \int_0^1
\nu^2(\tau) d \tau$, and $\nu(\tau) = \int_0^{\tau} \gamma(\tau')
d \tau'$. The averaging method is simplified with the gauge
transformation,
\begin{equation}
\label{final-transformation} w(x,t) = \psi(x,t) \exp \left( i
\nu_1 |\psi|^2(x,t)\right),
\end{equation}
which reduces (\ref{GP-averaged}) to the form:
\begin{eqnarray}
\nonumber i \psi_t - \nu_1 \psi |\psi|^2_t = & - & \psi_{xx} +
\gamma_0 |\psi|^2 \psi + V(x) \psi \\
\label{GP-final} & + & \mu \psi \left( |\psi|^2_{x} \right)^2,
\end{eqnarray}
where $\mu = \nu_2 - \nu_1^2$. Using the balance equation $i
|\psi|^2_t = \left( \bar{\psi}_x \psi - \bar{\psi} \psi_x
\right)_x$, which follows from (\ref{GP-final}), we rewrite 
the averaged equation in the final form:
\begin{eqnarray}
\label{GP-main}
\nonumber
i \psi_t = & - & \psi_{xx} + \gamma_0 |\psi|^2 \psi +
V(x) \psi + \mu \psi \left( |\psi|^2_{x} \right)^2 \\
& + & i \nu_1 \psi \left( \bar{\psi} \psi_x - \bar{\psi}_x \psi
\right)_x.
\end{eqnarray}

The averaged equation (\ref{GP-main}) is the main result of this
Letter. It is seen to be equivalent to the integral averaged equation derived
for strong dispersion management in fiber optics
\cite{DMS1,PZ}, but it is a {\it local} evolution equation. A
similar local equation was also derived for weak dispersion
management in fiber optics \cite{YK}, when the last two terms of
(\ref{GP-main}) are small compared to the leading-order NLS
equation. We emphasize that our main averaged equation
(\ref{GP-main}) is derived for strong nonlinearity management
and it captures all terms in the same, leading order of the
averaging method.

{\it Solitons in BECs}. The simplest standing waves 
of the averaged equation (\ref{GP-main}) are obtained 
through the standard ansatz \cite{sulem}:
\begin{equation}
\label{soliton} \psi(x,t) = \phi(x) e^{i \omega t},
\label{sw}
\end{equation}
where $\phi(x)$ solves the second-order differential equation:
\begin{equation}
\label{ODE} - \phi'' + \omega \phi + V(x) \phi + \gamma_0 \phi^3 +
4 \mu (\phi')^2 \phi^3 = 0.
\end{equation}

As a typical example of a smooth periodic variation of the
scattering length \cite{abdul,saito}, we use the sinusoidal
function $\Gamma(t)= \gamma_0 + \gamma_1 \sin(2 \pi t)$, in which
case $\mu=\gamma_1^2/(8 \pi^2)$. We also set $\epsilon=1$ and
choose $|\omega| \in [0.1,0.5]$ to ensure validity of the averaged
equation (\ref{ODE}), when $\epsilon \ll 2 \pi/|\omega|$. We also
use the parabolic potential $V(x)$ for the magnetic trapping of
the BEC, $V(x) = \frac{1}{2} \Omega^2 x^2$, where $\Omega^2 \in
[0.02,0.4]$.

To estimate actual physical quantities corresponding to the above
values of the normalized parameters, we first note that the cases
$\gamma_0 < 0$ ($\gamma_0 > 0$) are relevant to an attractive
(repulsive) condensate, such as $^{7}$Li ($^{85}$Rb),
characterized by a negative (positive) scattering length $a=-1$nm
($a=0.8$nm), in a magnetic field $B\approx 650$ G ($B\approx 159$
G). These values of the scattering lengths set the units in the
parameters $\gamma_0$ and $\gamma_1$, which may take different
values as long as the magnetic field $B$ is varied \cite{inouye}.
On the other hand, the number of atoms $N$ in the two condensates
is taken to be as follows: $N= 1 \times 10^4$ ($N=2 \times 10^5$)
for $\Omega^2=0.4$ ($\Omega^2=0.02$) for the $^{7}$Li condensate
and $N=4\times 10^3$ ($N=7.5 \times 10^4$) for $\Omega^2=0.4$
($\Omega^2=0.02$) for the $^{85}$Rb condensate. Since we deal with
cigar-shaped BECs, we may assume that the external magnetic trap
is highly anisotropic, characterized by the confining frequencies
$\omega_{\parallel}=2\pi \times 3.6$Hz and $\omega_{\perp}=2\pi
\times 360$Hz in the axial and transverse directions respectively.
In such a case, the time and space units in the results that will
be presented below, are $44.2$ms and $2\mu$m (for $^{7}$Li) or
$44.2$ms and $0.6\mu$m (for $^{85}$Rb).

{\it Numerical Results}. Using Eq. (\ref{ODE}), we can now obtain
the solution $\phi(x)$ for a given set of parameters
($\gamma_0,\gamma_1,\Omega,\omega$), by means of the Newton
method. We also perform parameter continuations, to follow the
solution branches as the parameters vary.

\begin{figure}
{\epsfig{file=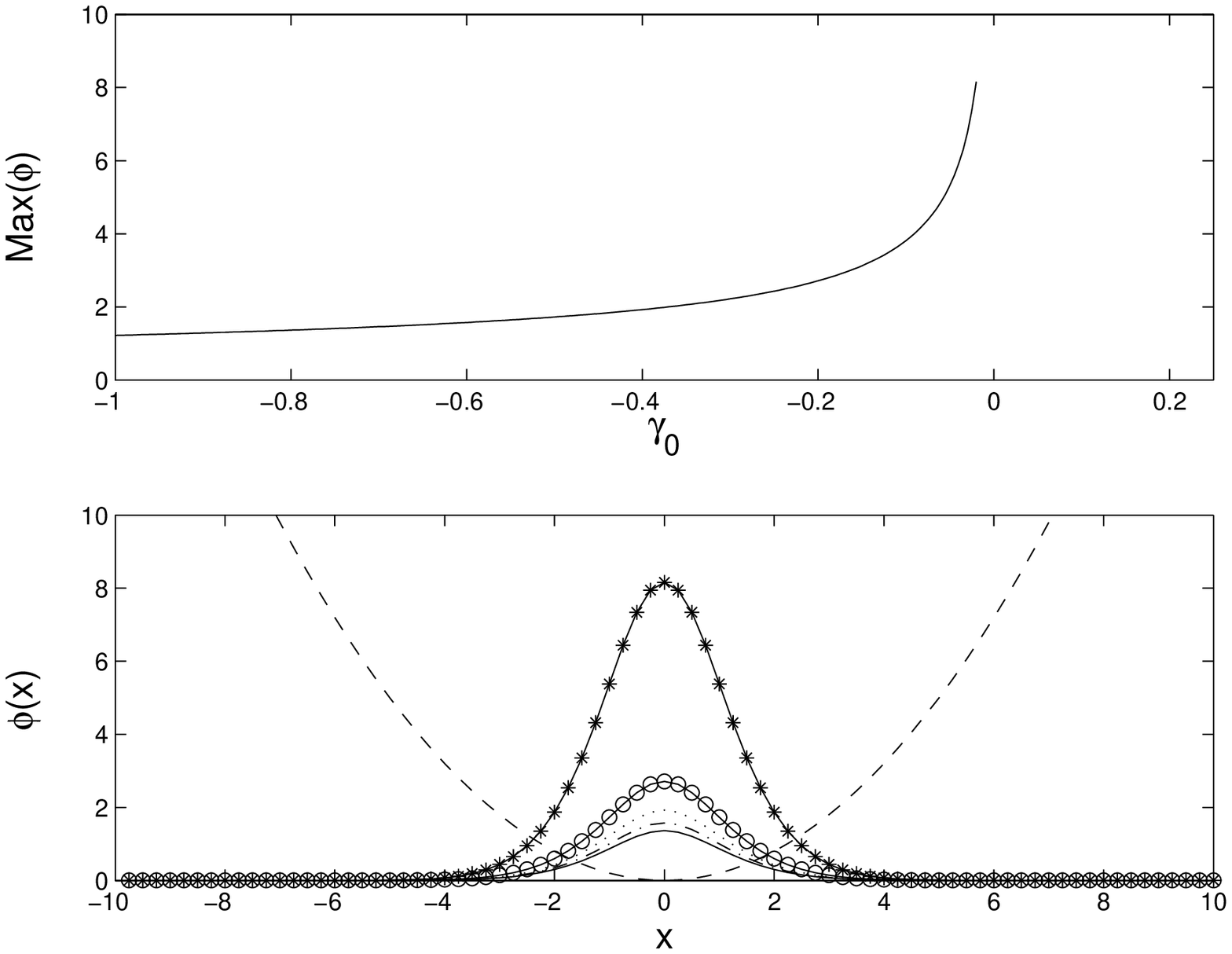, width=7.8cm,angle=0, clip=}}
{\epsfig{file=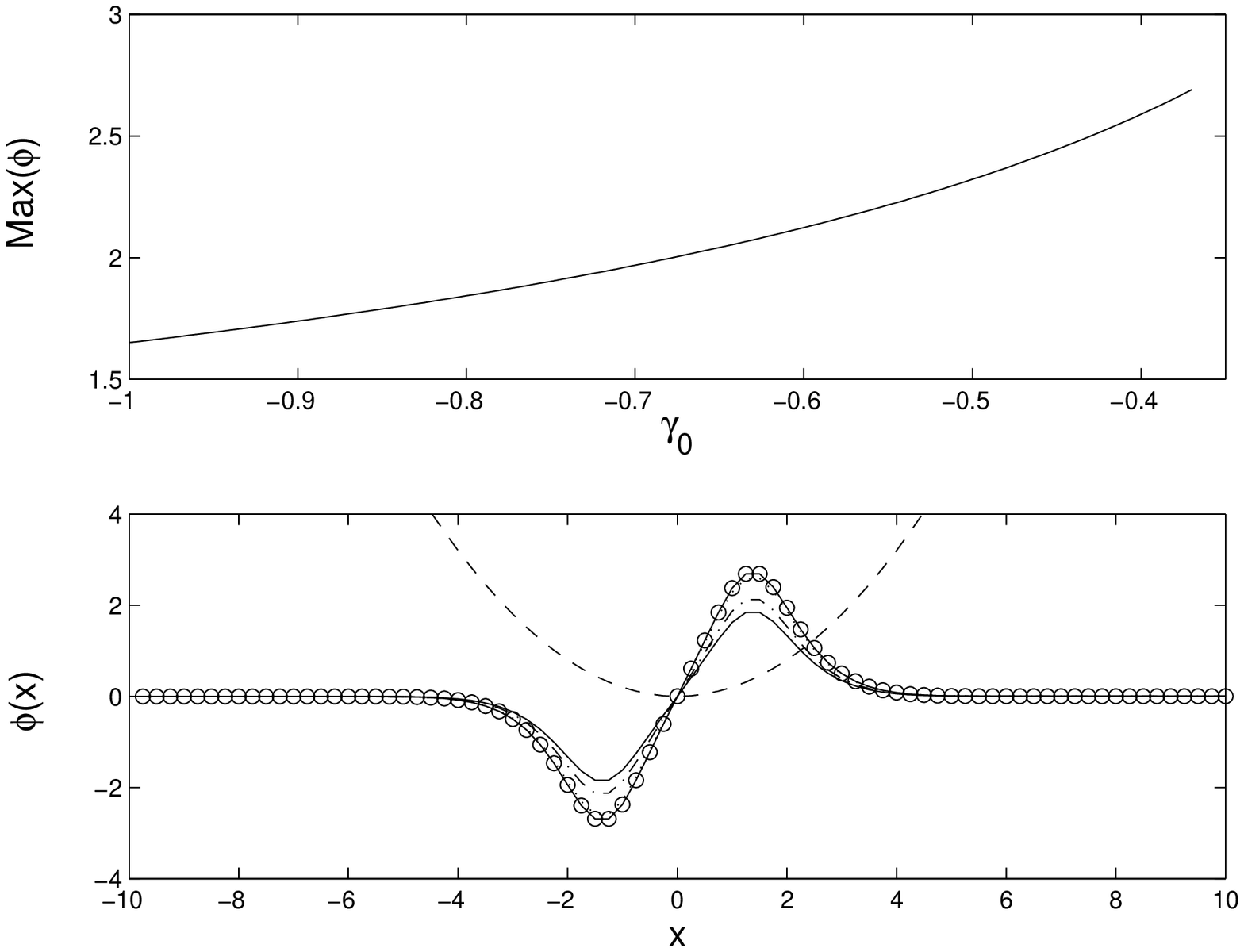, width=7.8cm,angle=0, clip=}}
\caption{Bright solitons (left) and twisted solitons (right)
obtained from Eq. (\ref{ODE}) with $\gamma_0 < 0$, $\gamma_1=0.5$,
$\Omega^2=0.4$, and $\omega = 0.5$. The top subplot shows the
solution maximum for different values of $\gamma_0$. The bottom
subplot shows the potential (dashed line) and the solutions: at
the left panel the solution is shown 
for $\gamma_0=-0.8$ (solid), $-0.6$ (dash-dotted),
$-0.4$ (dotted) $-0.2$ (circles) and $-0.02$ (stars) and at the
right panel for $\gamma_0=-0.8$ (solid), $-0.6$ (dash-dotted) and
$-0.37$ where the branch ceases to exist (circles).} \label{fign1}
\end{figure}

Fig. \ref{fign1} shows two solutions of the averaged equation
(\ref{ODE}) with $\gamma_0 < 0$ (the attractive BEC with negative
scattering length), $\gamma_1 = 0.5$, $\Omega^2 = 0.4$, and
$\omega = 0.5$. The solution on the left panel is the bright
soliton, which has the form $\phi(x) = (2 \omega/\gamma_0)^{1/2}
{\rm sech}(\omega^{1/2} x)$ when $\gamma_1 = \Omega = 0$. The
solution on the right panel is the so-called twisted soliton,
which corresponds to a concatenation of two separated bright
solitons of opposite parity (see e.g., \cite{TLM}). The twisted
soliton does not exist when $\gamma_0 < 0$ and $\Omega = 0$.
Higher-order solutions with multiple nodes (zeros) may also exist
in the averaged equation (\ref{ODE}) with $\gamma_0 < 0$ and
$\Omega \neq 0$, in some parameter domains.

\begin{figure}
{\epsfig{file=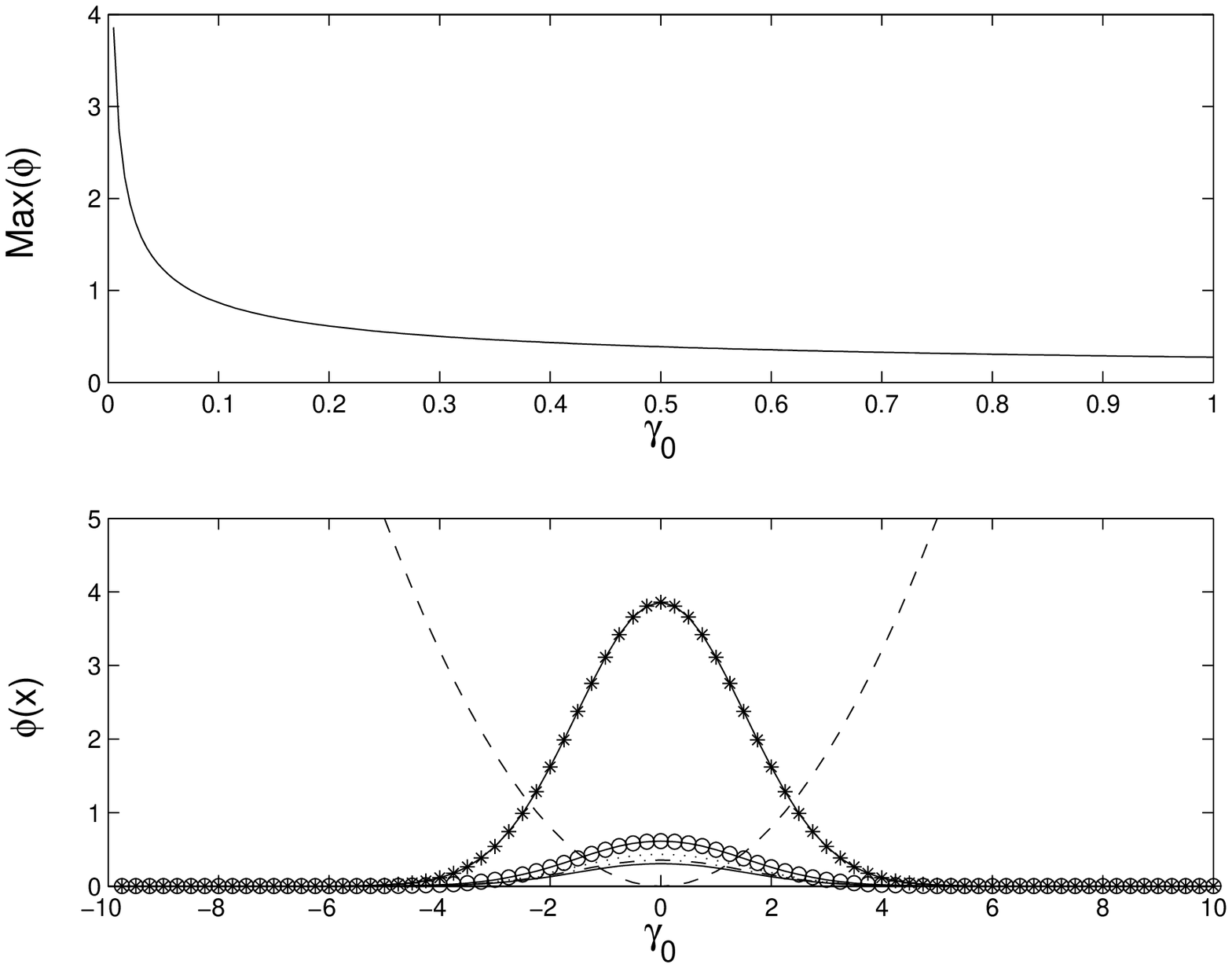, width=7.8cm,angle=0, clip=}}
{\epsfig{file=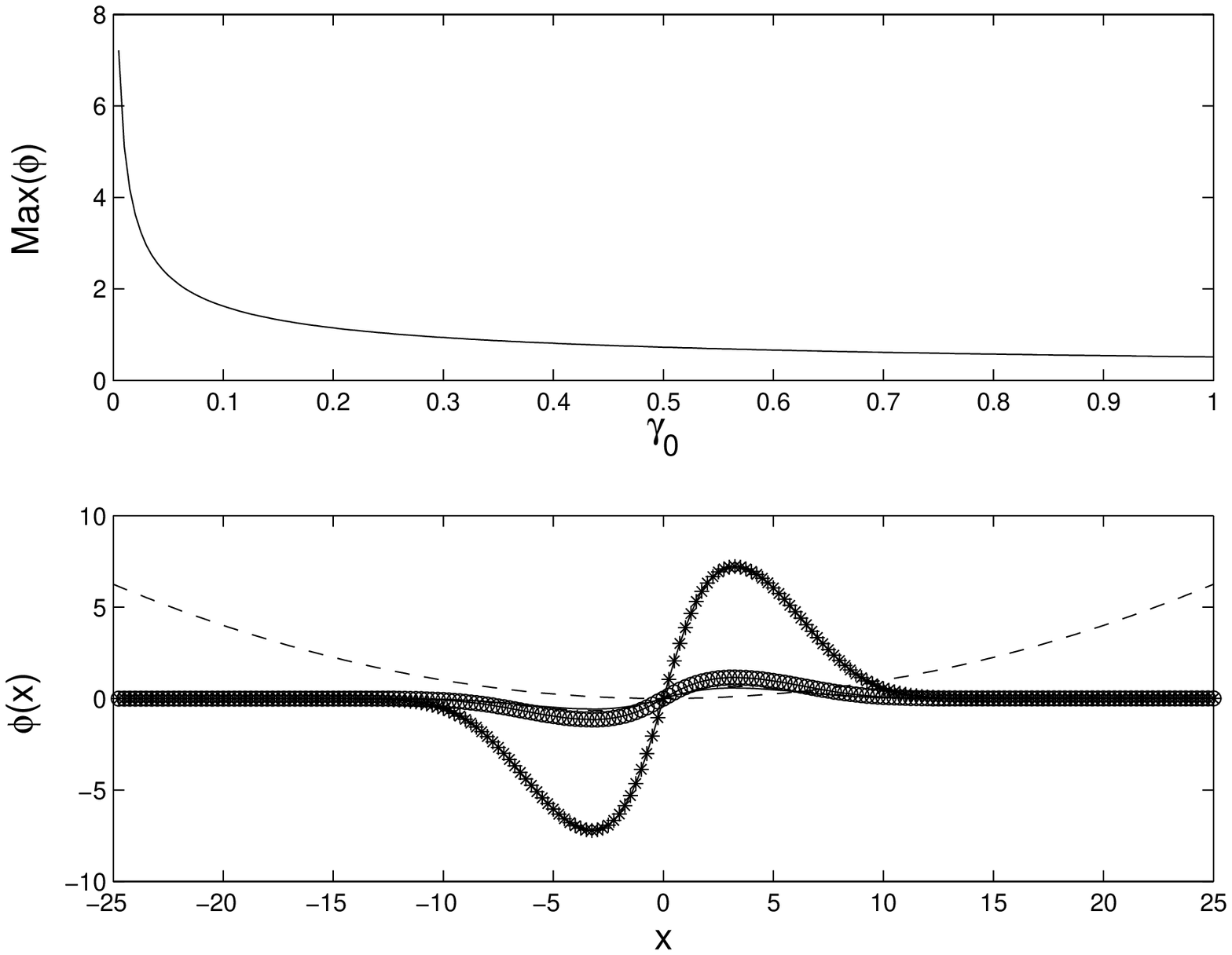, width=7.8cm,angle=0, clip=}}
\caption{Similar to the first figure but for 
Thomas-Fermi clouds (left) and dark solitons (right)
obtained from Eq. (\ref{ODE}) with $\gamma_0 > 0$, $\gamma_1=0.5$,
$\omega = 0.5$, while $\Omega^2 = 0.4$ (left) and $\Omega^2 =
0.02$ (right). The top subplot shows the solution maximum for
different values of $\gamma_0$. The bottom subplot shows the
potential (dashed line) and the solutions for $\gamma_0 = 0.8$
(solid) , $0.6$ (dash-dotted), $0.4$ (dotted), $0.2$ (circles) and
$0.01$ (stars). Notice that in the case of the dark soliton in
the right subplot, the solution profile is altered very slightly
between the cases  $\gamma_0=0.8$ and $\gamma_0=0.2$, which are
practically indistinguishable. } \label{fign1a}
\end{figure}

Fig. \ref{fign1a} shows two solutions of the averaged equation
(\ref{ODE}) with $\gamma_0 > 0$ (the repulsive BEC with positive
scattering length), $\gamma_1 = 0.5$, and $\omega = -0.5$. In the
case of $\gamma_0 > 0$, the localized solutions of Eq. (\ref{ODE})
bifurcate from linear modes trapped by the parabolic potential
$V(x)$, such that an infinite number of solitons with multiple
nodes (zeros) exists for larger negative values of $\omega$. The
solution on the left panel for $\Omega^2 = 0.4$ is the ground
state, often approximated by the Thomas-Fermi cloud \cite{KTFM}.
The solution on the right panel for $\Omega^2 = 0.02$ is the 
(embedded in the Thomas-Fermi cloud) dark
soliton  which, in the
case of  $\gamma_1 = \Omega = 0$,
has the form $\phi(x) = (|\omega|/\gamma_0)^{1/2}
{\rm tanh}((|\omega|/2)^{1/2} x)$ when $\gamma_1 = \Omega = 0$.
The dark soliton is the only localized solution of Eq. (\ref{ODE})
with $\gamma_0 > 0$ and $\Omega = 0$. Notice that the regular dark
soliton asymptotes to a non-vanishing amplitude when $\Omega=0$, while
it asymptotes to $0$ in the presence of the magnetic trap.

\begin{figure}
\centering{\epsfig{file=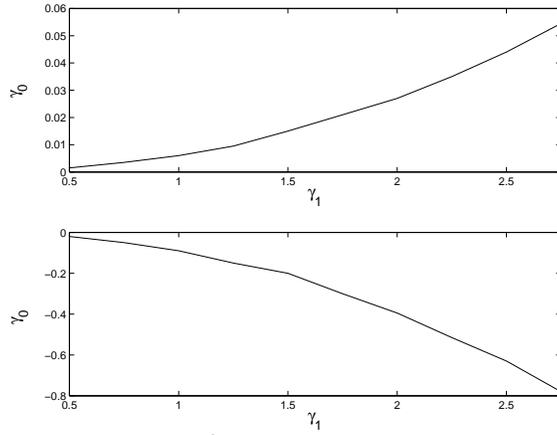, width=7.5cm,angle=0, clip=}}
\caption{Domain of existence of dark (top panel for
$\Omega^2=0.02$) and bright (bottom panel for $\Omega^2=0.4$)
solitons of Eq. (\ref{ODE}) with $|\omega| = 0.5$. The solutions
exist above and below, respectively, the corresponding curves of
the ($\gamma_0,\gamma_1$) plane.} \label{fign2}
\end{figure}

Fig. \ref{fign2} shows two parameter $(\gamma_0,\gamma_1)$
continuation of the dark (top panel for $\Omega^2=0.02$) and
bright (bottom panel for $\Omega^2=0.4$) soliton solutions of the
averaged equation (\ref{ODE}) with $|\omega| = 0.5$. The branch of
dark solitons exists above the bifurcation curve on the top panel,
whereas the branch of bright solitons exists below the curve on
the bottom panel. The two curves pass through the origin $\gamma_0 =
\gamma_1 = 0$. We note that the domain of existence of dark and
bright solitons shrinks for increasing values of $\gamma_1$.

Finally, we examine how well the averaged equation (\ref{ODE})
approximates bright and dark solitons of the GP equation
(\ref{GP}). In our numerical simulations of Eq. (\ref{GP}), we
initialize the wavefunction, using the spatial profile obtained
from (\ref{ODE}), and subsequently observe whether the temporal
evolution of Eq. (\ref{GP}) preserves the average profile of Eq.
(\ref{ODE}).

\begin{figure}
\centering {\epsfig{file=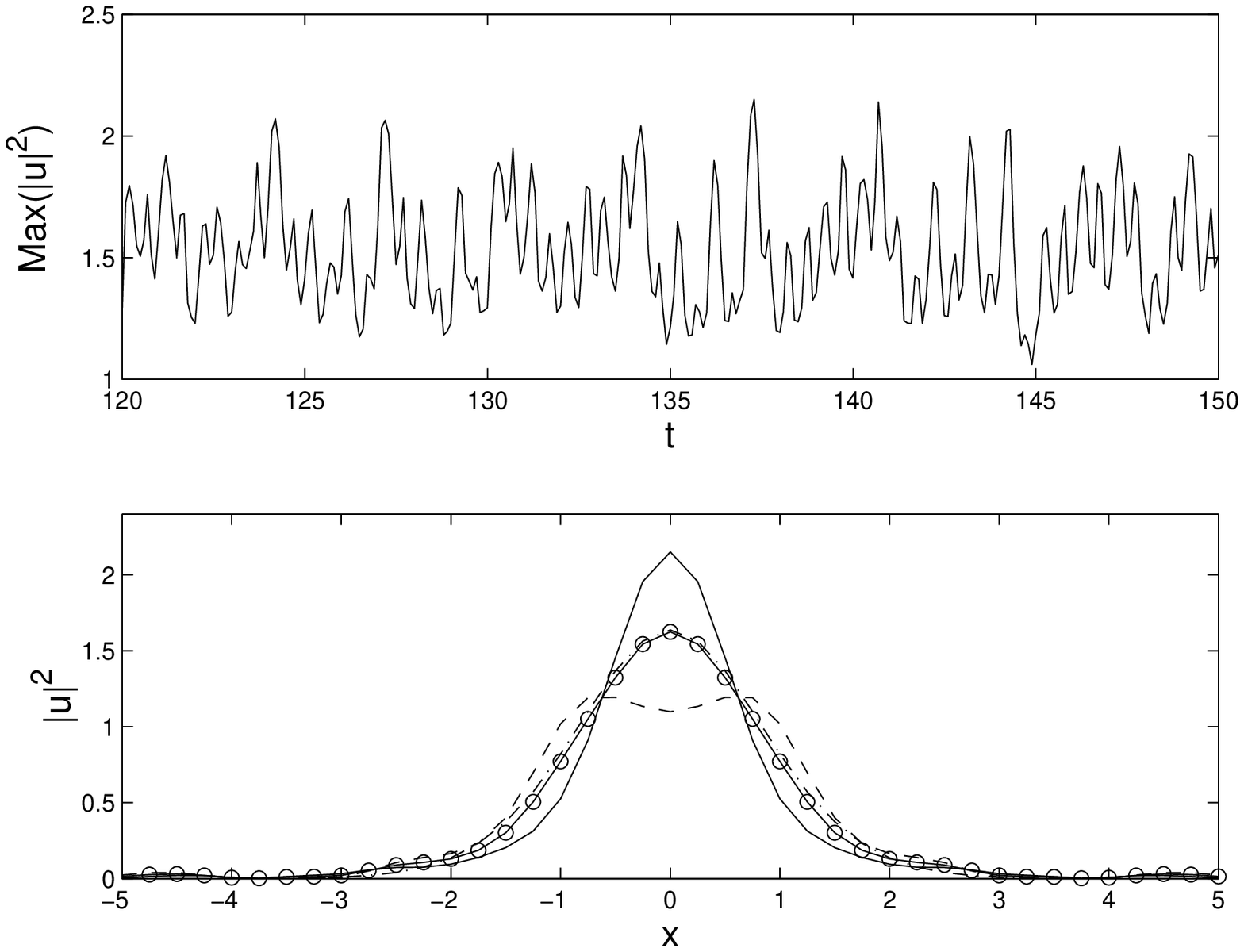, width=9cm,angle=0, clip=}}
{\epsfig{file=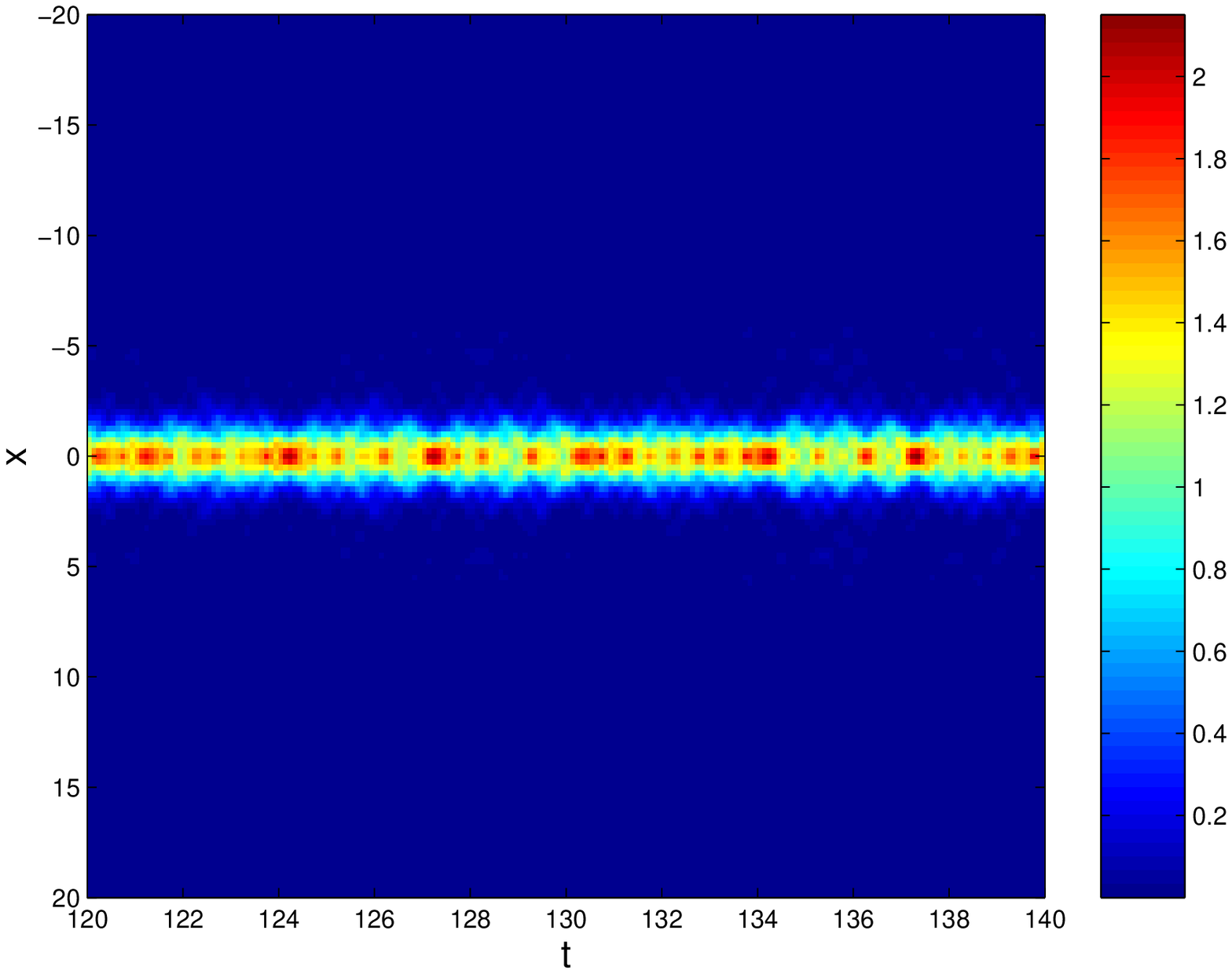, width=9cm,angle=0, clip=}}
\caption{Temporal evolution of the bright soliton with
$\gamma_0=-0.5$, $\gamma_1=1$, $\Omega^2=0.4$, and $\omega = 0.5$.
The top panel shows evolution of the (spatial) maximum of
$|u|^2(x,t)$ as a function of time. The middle panel shows the
solution $|u|^2(x,t)$ at $t=137.3$ (solid) and $t=138$ (dashed),
their average (circles), and the initial configuration
(dash-dotted). The latter practically {\it coincides} with
the average. The bottom panel shows a contour plot of
$|u|^2(x,t)$ in $(x,t)$.} \label{fign3}
\end{figure}

Fig. \ref{fign3} shows the temporal evolution of the bright
soliton with $\gamma_0=-0.5$, $\gamma_1=1$, $\Omega^2 = 0.4$, and
$\omega = 0.5$. The periodic variations of the nonlinearity
coefficient $\Gamma(t)$ in Eq. (\ref{GP}) lead to complicated
oscillations of the solution's maximum. While the solution
oscillates between single-humped and double-humped solitons (a
scenario that bears analogies to the observations of \cite{KTFM}),
the average of the two extreme solitons (at the maximum and the
minimum amplitudes) is practically {\it indistinguishable} from the profile
$\phi(x)$ of Eq. (\ref{ODE}).

\begin{figure}
\centering {\epsfig{file=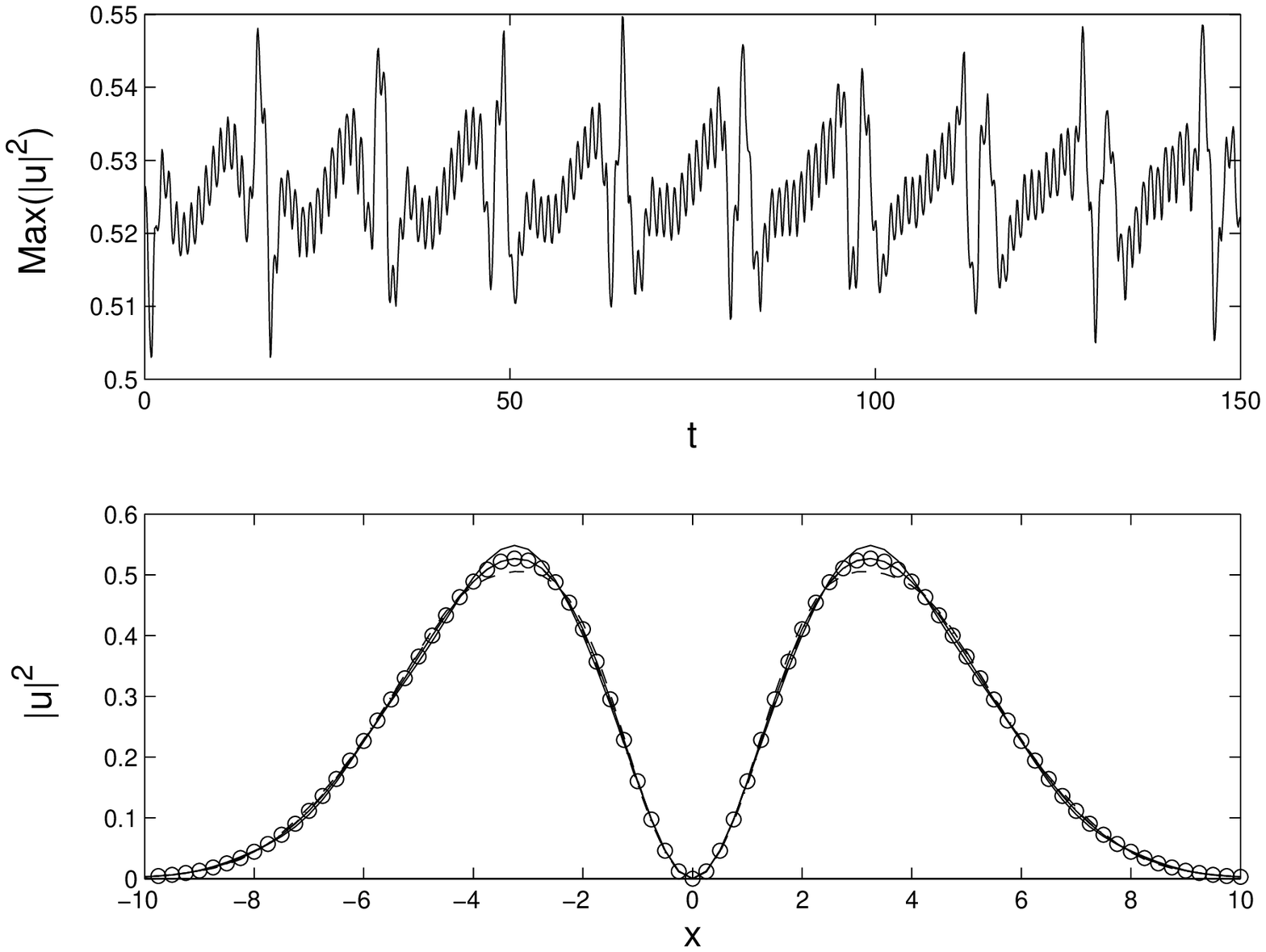, width=9cm,angle=0, clip=}}
{\epsfig{file=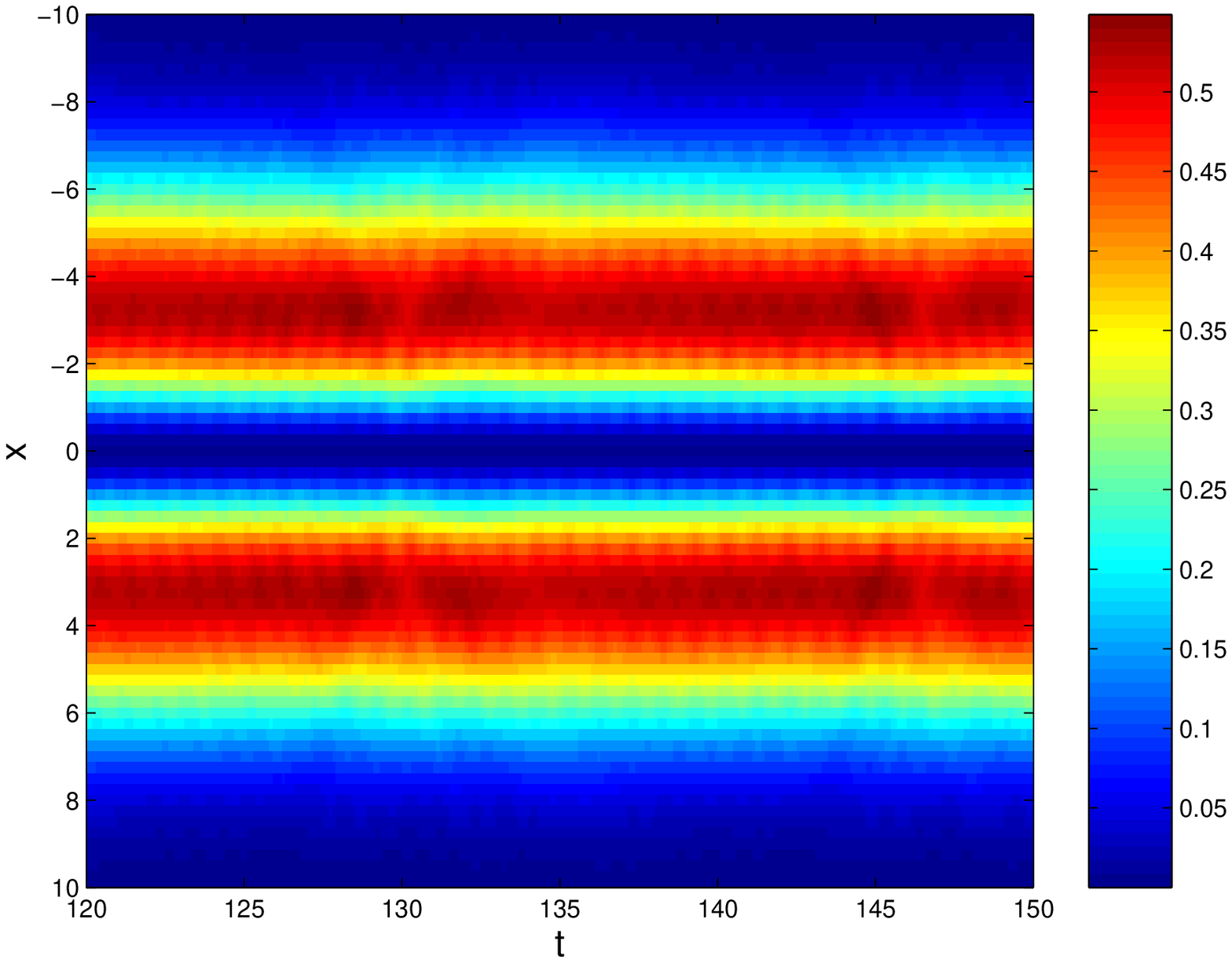, width=9cm,angle=0, clip=}} \caption{Same as
Fig. \ref{fign3}, but for the dark soliton with $\gamma_0=0.5$,
$\gamma_1=1$, $\Omega^2=0.02$, and $\omega = -0.5$. The maximum
and minimum snapshots correspond to $t=128.4$ and $t=130.2$.
The dash-dotted line of the theoretical prediction (initial condition)
again {\it coincides} with the average denoted by the circles.}
\label{fign4}
\end{figure}

Fig. \ref{fign4} shows the temporal evolution of the dark soliton
with $\gamma_0=0.5$, $\gamma_1=1$, $\Omega^2 = 0.02$, and $\omega
= -0.5$. We notice that the center of the dark soliton remains at
the origin $x = 0$, without any oscillations. Only the maxima of
$|u|^2(x,t)$ display periodic oscillations of small amplitude. The
average of the extreme solitons is again essentially identical to the profile
$\phi(x)$ of the averaged equation (\ref{ODE}).

{\it Conclusion}. We have derived and studied the averaged
equation (\ref{GP-main}) for the NLS (GP) equation (\ref{GP}) with
periodic modulation of the nonlinearity coefficient. Our results 
are of broad interest to diverse areas of atomic and optical physics, 
as well as of nonlinear and,  also, mathematical
physics. We have
identified numerically several branches of solitary waves of
the averaged stationary equation (\ref{ODE}). We have also
compared solutions of the averaged and full equations, obtaining a
very good agreement. It is of particular and immediate interest to examine 
these predictions experimentally, as well as to identify what happens
in parametric regimes where the present theory is no longer able
to identify such waves. Furthermore,
the averaged equation (\ref{GP-main}) can be
useful for future studies of BECs, e.g., in
optical lattice potentials or in higher dimensions.

This work was supported in part by NSERC grant \# RGP238931 (DEP),
and by a UMass FRG, NSF-DMS-0204585 and the Eppley Foundation for
Research (PGK).

\end{multicols}

\begin{thebibliography}{200}
\bibitem{sulem} C. Sulem and P.L. Sulem, \newblock
{\it The Nonlinear Schr{\"o}dinger Equation},
Springer-Verlag (New York, 1999).

\bibitem{sulem1} For a recent survey of optical applications see
Yu.S. Kivshar and G.P. Agrawal, {\it Optical Solitons From
Fibers to Photonic Crystals}, Academic Press (2003).

\bibitem{sulem2} For a recent survey of atomic physics applications
see F. Dalfovo {\it et al.}, Rev. Mod. Phys. {\bf 71}, 463 (1999).

\bibitem{DMS} For a recent survey of dispersion management, see
S.K. Turitsyn {\it et al}, C.R. Physique {\bf 4}, 145 (2003).

\bibitem{DMS1} I. Gabitov and S. Turitsyn, Opt. Lett. {\bf 21}, 327 (1996);
I. Gabitov and S. Turitsyn, JETP Lett. {\bf 63}, 814 (1996); M.J.
Ablowitz and G. Biondini, Opt. Lett. {\bf 23}, 1668 (1998).

\bibitem{PZ} D.E. Pelinovsky and V. Zharnitsky, SIAM J. Appl.
Math. {\bf 63}, 745 (2003).

\bibitem{YK} T.S. Yang and W.L. Kath, Opt. Lett. {\bf 22}, 985
(1997); T.I. Lakoba and D.E. Pelinovsky, Chaos {\bf 10}, 539
(2000).

\bibitem{mal1} I. Towers and B.A. Malomed, J. Opt. Soc. Am. {\bf 19}, 537
(2002); L. Berg{\'e} et al., Opt.  Lett. {\bf 25}, 1037 (2000).

\bibitem{inouye}  S. Inouye {\it et al.}, Nature {\bf 392}, 151 (1998);
J. Stenger {\it et al.}, Phys. Rev. Lett. {\bf 82}, 2422 (1999);
J.L. Roberts {\it et al.}, Phys. Rev. Lett. {\bf 81}, 5109 (1998);
S.L. Cornish {\it et al.}, Phys. Rev. Lett. {\bf 85}, 1795 (2000);
E.A. Donley {\it et al.}, Nature {\bf 412}, 295 (2001).

\bibitem{KTFM} P.G. Kevrekidis {\it et al.},
Phys. Rev. Lett. {\bf 90}, 230401 (2003).

\bibitem{abdul} F.Kh. Abdullaev {\it et al.},
Phys. Rev. Lett. {\bf 90}, 230402 (2003); F.Kh. Abdullaev {\it et
al.}, Phys. Rev. A {\bf 67}, 013605 (2003); F.Kh. Abdullaev {\it
et al.}, cond-mat/0306281.

\bibitem{saito} H. Saito and M. Ueda, Phys. Rev. Lett. {\bf 90},
040403 (2003).

\bibitem{solitons} S. Burger {\it et al.}, Phys. Rev. Lett. {\bf 83},
5198(1999); J. Denschlag {\it et al.}, Science {\bf 287}, 97
(2000); B. P. Anderson {\it et al.}, Phys. Rev. Lett. {\bf 86},
2926 (2001); K.E. Strecker {\em et al.}, Nature {\bf 417}, 150
(2002); L. Khaykovich {\em et al.}, Science {\bf 296}, 1290
(2002).

\bibitem{herrero} V.M. P\'{e}rez-Garc\'{i}a, H. Michinel and H. Herrero,
Phys. Rev. A {\bf 57}, 3837 (1998); L. Salasnich, A. Parola and L.
Reatto, Phys. Rev. A {\bf 65}, 043614 (2002);
Y.B. Band, I. Towers, and B.A. Malomed, Phys. Rev. A 67, 023602 (2003).

\bibitem{sanders} J.A. Sanders and F. Verhulst,
{\it Averaging Methods in Nonlinear Dynamical Systems}, Springer-Verlag
(Berlin, 1985).

\bibitem{TLM} P.G. Kevrekidis {\it et al.}, New J. Phys. {\bf 5}, 64 (2003).

\end{thebibliography}
\end{document}